\documentclass[acmsmall,nonacm]{acmart}
\usepackage{amsmath,amsthm,amsbsy}
\usepackage{microtype}
\usepackage[ruled]{algorithm2e}
\usepackage{setspace}
\usepackage{color}
\usepackage{xcolor}
\usepackage[noend]{algpseudocode}
\usepackage{multirow}
\usepackage{paralist}
\usepackage{xspace}

\newcommand{\ra}{\rightarrow}

\renewcommand{\Pr}{\operatorname*{\textbf{\textup{Pr}}}}
\DeclareMathOperator*{\II}{\textbf{\textup{I}}}

\DeclareMathOperator*{\HH}{\textbf{\textup{H}}}

\DeclareMathOperator*{\EE}{\textbf{\textup{E}}}

\renewcommand{\geq}{\geqslant}
\renewcommand{\ge}{\geqslant}
\renewcommand{\leq}{\leqslant}
\renewcommand{\le}{\leqslant}
\newcommand{\set}[1]{\{ #1 \}}

\DeclareMathOperator{\poly}{\ensuremath{\mathrm{poly}}}

\newcommand{\congest}{\ensuremath{\mathsf{CONGEST}}}

\makeatletter
\newtheorem*{rep@theorem}{\rep@title}
\newcommand{\newreptheorem}[2]{%
\newenvironment{rep#1}[1]{%
\def\rep@title{#2 \ref{##1}}%
\begin{rep@theorem}[restated]}%
{\end{rep@theorem}}}
\makeatother
\newreptheorem{lemma}{Lemma}
\newreptheorem{theorem}{Theorem}

\newboolean{short}
\setboolean{short}{false}
\newcommand{\onlyShort}[1]{\ifthenelse{\boolean{short}}{#1}{}}
\newcommand{\onlyLong}[1]{\ifthenelse{\boolean{short}}{}{#1}}

 \theoremstyle{definition}
\newtheorem{problem}{Problem}
 \newtheorem{claim}{Claim}
 \newtheorem{corollary}{Corollary}
\newtheorem{open_problem}{Open Problem}
\newtheorem{theorem}{Theorem}
\newtheorem{fact}{Fact}

\usepackage[most]{tcolorbox}
\tcbuselibrary{theorems}
\newtcbtheorem[number within=section]{Definition}{Definition}{
  enhanced,
  sharp corners,
  attach boxed title to top left={
    yshifttext=-1mm
  },
  colback=white,
  colframe=gray!75!white,
  fonttitle=\bfseries,
  boxed title style={
    sharp corners,
    size=small,
    colback=gray!75!white,
    colframe=gray!75!white,
  }
}{def}

\tcolorboxenvironment{reptheorem}{lower separated=false,
                colback=white!90!gray,
colframe=white, fonttitle=\bfseries,\usepackage{amsmath,amsthm,amsbsy,amssymb}
\usepackage{microtype}
\usepackage{setspace}
\usepackage{color}
\usepackage{xcolor}
\usepackage{algorithm}
\usepackage[noend]{algpseudocode}
\usepackage{multirow}
\usepackage{xspace}
\usepackage{graphicx}
\usepackage{subcaption}

\let\originalleft\left
\let\originalright\right
\renewcommand{\left}{\mathopen{}\mathclose\bgroup\originalleft}
\renewcommand{\right}{\aftergroup\egroup\originalright}

\newcommand{\ra}{\rightarrow}

\renewcommand{\Pr}{\operatorname*{\textbf{\textup{Pr}}}}
\DeclareMathOperator*{\II}{\textbf{\textup{I}}}

\DeclareMathOperator*{\HH}{\textbf{\textup{H}}}

\DeclareMathOperator*{\EE}{\textbf{\textup{E}}}

\renewcommand{\geq}{\geqslant}
\renewcommand{\ge}{\geqslant}
\renewcommand{\leq}{\leqslant}
\renewcommand{\le}{\leqslant}
\newcommand{\set}[1]{\{ #1 \}}

\newcommand{\lt}{\left}
\newcommand{\rt}{\right}
\newcommand{\md}{\middle}

\DeclareMathOperator{\poly}{\ensuremath{\mathrm{poly}}}

\newcommand{\congest}{\ensuremath{\mathsf{CONGEST}}}

\makeatletter
\newtheorem*{rep@theorem}{\rep@title}
\newcommand{\newreptheorem}[2]{%
\newenvironment{rep#1}[1]{%
\def\rep@title{#2 \ref{##1}}%
\begin{rep@theorem}[restated]}%
{\end{rep@theorem}}}
\makeatother
\newreptheorem{lemma}{Lemma}
\newreptheorem{theorem}{Theorem}

\newtheorem{lemma}{Lemma}

\newtheorem{theorem}{Theorem}
\theoremstyle{plain}

\newtheorem{invariant}{Invariant}     

\usepackage[most]{tcolorbox}
\tcbuselibrary{theorems}
\newtcbtheorem[number within=section]{Definition}{Definition}{
  enhanced,
  sharp corners,
  attach boxed title to top left={
    yshifttext=-1mm
  },
  colback=white,
  colframe=gray!75!white,
  fonttitle=\bfseries,
  boxed title style={
    sharp corners,
    size=small,
    colback=gray!75!white,
    colframe=gray!75!white,
  }
}{def}

colbacktitle=white!50!gray,
coltitle=black,
enhanced,
boxed title style={colframe=black},
}

\usepackage{thm-restate}
\usepackage{color,soul}   
\usepackage{enumitem}
\usepackage{hyperref}
\hypersetup{
     colorlinks=true,
     linkcolor=blue,
     citecolor = blue,      
     urlcolor=cyan,
}
\usepackage{subcaption}

\usepackage{geometry}
\newcommand{\spreading}{\textsf{Spreading}}    
\newcommand{\EdgeDel}{\textsf{QuickBipartiteMM}}    
\newtheorem{result}{Result}

\newcommand{\lt}{\left}
\newcommand{\rt}{\right}
\newcommand{\md}{\middle}
\newcommand{\kclique}{k\text{-}\mathsf{CLIQUE}_\beta }


\author{Peter Robinson}
\affiliation{
  \department{School of Computer \& Cyber Sciences}
  \institution{Augusta University}
  \country{USA}
}

\author{Xianbin Zhu}
\affiliation{
  \department{Department of Computer Science}
  \institution{Aalto University}
  \country{Finland}
}

\setboolean{short}{false}
\begin{document}
\title{Approximate Maximum Matching in the Distributed Vertex Partition Model under Dynamic Graph Updates}
\begin{abstract}
We initiate the study of approximate maximum matching in the vertex partition model, for graphs subject to dynamic changes. 
We assume that the $n$ vertices of the graph are partitioned among $k$ players, who execute a distributed algorithm and communicate via message passing.
An adaptive adversary may perform dynamic updates to the graph topology by inserting or removing edges between the nodes, and the algorithm needs to respond to these changes by adapting the output of the players, with the goal of maintaining an approximate maximum matching. 
The main performance metric in this setting is the algorithm's update time, which corresponds to the number of rounds required for updating the solution upon an adversarial change. 

For the standard setting of single-edge insertions and deletions, we give a randomized Las Vegas algorithm with an expected update time of $O\lt( \lceil \frac{\sqrt{m}}{\beta k} \rceil \rt)$ rounds that maintains a $\frac{2}{3}$-approximate maximum matching that is also maximal, where $m$ is the number of edges in the graph and $\beta$ is the available link bandwidth.
For batch-dynamic updates, where the adversary may insert up to $\ell\ge 1$ edges at once, we prove the following:
\begin{itemize} 
\item There is a randomized algorithm that succeeds with high probability in maintaining a $\frac{2}{3}$-approximate maximum matching and has a worst case update time of $O\lt( \lt\lceil\frac{\ell\log n}{\sqrt{\beta k}}\rt\rceil \rt)$ rounds.
\item Any algorithm for maintaining a maximal matching without 3-augmenting paths under batches of $\ell$-edge insertions has an update time of $\Omega\lt( \frac{\ell}{\beta k \log n} \rt)$ rounds in the worst case.
\end{itemize}

\end{abstract}
\maketitle


\section{Introduction}

Modern-day computing systems produce large amounts of data at an astonishing rate.
These data sets often form complex relationships that are best represented as a graph consisting of vertices that are connected by edges.
Performing efficient computations on such massive graphs is a fundamental challenge encountered in various applications, ranging from machine learning to protein interaction graphs and social network analysis.
To cope with the scale of such massive graphs, which often exceeds the memory and bandwidth capabilities of a single machine, several iterative graph processing frameworks such as Google Pregel~\cite{pregel}, and various open source projects such as, e.g., 
Apache Giraph~\cite{facebookgiraph}, have been developed and are frequently used in industrial applications of graph analytics.
In contrast to systems based on the MapReduce~\cite{dean2008mapreduce} paradigm, these frameworks follow the \emph{vertex-centric} programming model~\cite{pregel}, where developers express their algorithms by ``thinking like a vertex.'' Intuitively speaking, this means that each vertex contains information about itself and its incident edges, and the computation is expressed at the level of a single vertex, which is very similar in spirit to traditional distributed graph algorithms~\cite{peleg_book}, and assumes that the input graph is \emph{vertex-partitioned} among the physical machines forming the network. 
Moreover, some frameworks such as Giraph++~\cite{tian2013think} and Blogel~\cite{yan2014blogel} expose a \emph{subgraph-centric} programming interface in addition to the vertex-centric one, which provides greater flexibility by allowing algorithms that take into account the state of a subset of vertices as part of their state transition function. 



\subsection{The $k$-Clique Message Passing Model} \label{sec:kclique}
When considering distributed graph algorithms in the big data setting, the size of the input graph typically dwarfs the number of available physical machines and the communication links between them. 
This is the underlying assumption of the \emph{$k$-clique message passing model}~\cite{li2024dynamic,amg} (denoted by $\kclique$), where $k$ players $P_1,\dots,P_k$ are interconnected using a clique topology of $\Theta(k^2)$ point-to-point links with a per-link bandwidth of $\beta$ bits. 
In more detail, the $\kclique$ model assumes that the input graph $G$ consists of $n \ge k$ vertices, $m$ edges, and $G$ is vertex-partitioned among the $k$ players in a roughly balanced way, such that each player obtains at most $O\lt(\frac{n\log n}{k}\rt)$ vertices.
We assume a separation between the unit that stores  (dynamic) input/output and the unit that performs computation which we call memory. All other data except input and output are stored in the memory. 

The players communicate via message passing over the bidirectional links of the clique network, and we assume that the computation proceeds in \emph{synchronous rounds}: 
In every round, each player performs some local computation, which may include private random coin flips, and then sends a message of at most $O(\beta\log n)$ bits over each one of its $k-1$ incident communication links, where $\beta \ge 1$ is the \emph{bandwidth parameter} of the model; note that $\beta$ may be $\omega(1)$.
All messages are guaranteed to be received by the end of the same round, and thus, in each round, every player can receive up to $O(\beta\,k\log n)$ bits in total from the other players. 

As mentioned, we focus on {vertex-partitioning}, which means that every vertex $u$ is \emph{assigned} to some player $P$ together with its incident edges; consequently, every edge of the input graph is known to two players initially.
Formally speaking, each vertex has a unique ID of $\Theta(\log n)$ bits, and ``knowing'' a vertex just means that this vertex's ID is part of the local state of $P$.
For each vertex $v$ assigned to $P$, we say that $P$ \emph{hosts} vertex $v$, and we use $P_u$ to refer to the player hosting vertex $u$. 
The input of a player $P$ consists of the IDs of its hosted vertices and, for each of its vertices, $P$ knows the IDs of $v$'s neighbors as well as the player to which they were assigned to. 
In this work, we consider an \emph{adaptive adversary}, who can choose any \emph{balanced vertex partitioning}, which requires each player to hold at most $O\lt( \frac{n}{k}\log n \rt)$ vertices in total.\footnote{Note that this is a main difference between the $\kclique$ model and the $k$-machine model~\cite{klauck2014distributed}, where the vertices are \emph{randomly} partitioned.} 

We point out that, for $k=n$ players and $\beta=1$, our model is equivalent to the congested clique~\cite{lotker} if each player obtains exactly one vertex.
Analogously to the congested clique and the $\congest$ model~\cite{peleg_book} and in contrast to the popular Massively Parallel Computation (MPC) model~\cite{mpc}, 
we do not impose a restriction on the local memory of the players in the $\kclique$ model.
That said, our results for single updates hold even for \emph{memoryless} dynamic algorithms that do not use any additional memory apart from storing the input and output. 
  
\subsection{Dynamic Updates}
To address the reality of evolving data sets in the real-world, we assume that the initial graph consists of $n$ vertices without any edges, and model dynamical changes by considering a sequence of updates to the edges of the graph. 
There are two important models in this setting: 
single-updates where each update is just the addition or the deletions of a single edge, and batch-dynamic updates, where each update consists of a \emph{batch} of at most $\ell$ edge deletions/insertions.\footnote{Batch-dynamic algorithms have attracted significant attention recently, see \cite{acar2019parallel,acar2011parallelism,anderson2023parallel,dhulipala2020parallel,dhulipala2021parallel,liu2022parallel,nowicki2021dynamic,gilbert2020fast,li2024dynamic}.}
In either case, the adaptive adversary may observe the local memory of the players, which includes the currently computed output, before choosing the next update. 
Since we are assuming a distributed algorithm, each player only needs to output the part of the solution relevant to its hosted vertices. 
That is, for computing a matching $M$, each player $P$ needs to output every edge $\set{u,v} \in M$, for which it hosts an endpoint.

Upon an update, every player that hosts a vertex $v$ incident to an added or deleted edge learns about $v$'s new neighborhood.
These changes yield a modified graph $G'$. 
The algorithm must then react by updating its current output to yield a solution for $G'$; no further changes to the graph topologies occur until the algorithm has terminated.
Subsequently, the next update arrives, and so forth.
We say that an \emph{algorithm has an update time of $T$}, if the algorithm outputs a valid solution following an update in at most $T$ rounds. 
Here, we assume that the local states of the players corresponded to a valid solution on the graph before the arrival of the latest updates. 

\medskip\noindent\textbf{Approximate Maximum Matching.}
In this work, we focus on computing a matching, which is one of the most fundamental problems in computer science. A \emph{matching} in a graph $G$ is a set of edges without common endpoints, and we say that a matching $M$ is a maximum (cardinality) matching if no matching in $G$ is larger than $M$. A matching $M$ is an \emph{$\alpha$-approximate matching} if $|M| \geq \alpha |\textsc{mcm}(G)|$ where  $|\textsc{mcm}(G)|$ is the cardinality of the maximum matching. It is well-known that the obtained matching is at least a $\frac{k}{k+1}$-approximation, if there is no augmenting paths of length $2k-1$ or less, see \cite{hopcroft1973n}, which we leverage in our algorithms: 
 

\begin{lemma}[\cite{hopcroft1973n}] \label{lem:hopcroft}
    Suppose a matching $M$ in a graph $G$ does not admit augmenting paths of length $2k-1$ or less. Then we have $|M| \geq \frac{k}{k+1} |M^*|$ where $M^*$ is the maximum matching in $G$.
\end{lemma}

\subsection{Contributions and Technical Challenges} \label{sec:contributions}

We present the first results for approximate maximum matching under dynamic updates against an adaptive adversary in the vertex partition message passing model. 
We first consider  $\kclique$ model under the standard setting where the adversary inserts or deletes a single edge at a time.
Here, we focus on algorithms that are memoryless, which means that the players do not keep any additional state across updates, apart from their input and output.

\begin{result}[informal]\label{res:1}
Maintaining a $\frac{2}{3}$-approximate maximum matching with a memoryless algorithm under single-edge insertions and deletions is possible with an update time of $ O(\lceil \frac{\sqrt{m}}{k\beta }\rceil )$ rounds, where $m$ is the current number of edges in the graph. 
\end{result}
For obtaining Result~\ref{res:1}, we show an edge-insertion can be dealt with by simply coordinating between the players that hold a vertex adjacent to the new edge and checking whether adding this edge to the matching induces a 3-augmenting path in the graph.

Handling an edge-deletion is more challenging, however, since a deleted matched edge may cause many nodes to become ``free'' (i.e., unmatched), which in turn may result in 3-augmenting paths that are difficult to find in sufficiently dense graphs.
More concretely, suppose that the matched edge $\set{u,v}$ was deleted and $u$ (1) has high node degree and (2) no free neighbor. 
We show that we can sample some neighbor $w$ of $u$ that itself has low degree, remove $w$'s edge from the matching and instead add the edge $\set{u,w}$.
As this operation may result in new 3-augmenting paths, we leverage the fact that $w$ has low degree, which allows us use an information dissemination routine of \cite{li2024dynamic} to find and resolve such a 3-augmenting path if it indeed exists.

Next, we turn our attention to the batch-dynamic model, where the adversary can perform up to $\ell$ updates instantaneously. 
As described above, handling even just a single edge-deletion requires a non-negligible amount of communication, and thus we exclusively focus on the batch-incremental setting, where the adversary can insert batches of edges. 
Our main technical contribution is an algorithm that handles batches of edge-insertions efficiently:

\begin{result}[informal]\label{res:2}
There is an algorithm that maintains a $\frac{2}{3}$-approximate maximum matching with an update time of $\widetilde O(\frac{\ell}{\sqrt{\beta k}})$ rounds, assuming that the adaptive adversary inserts a batch of $\ell$ edges.
In fact, any algorithm that maintains a maximal matching without 3-augmenting paths must have a worst case update time of at least $\widetilde \Omega(\frac{\ell}{\beta k})$ rounds.
\end{result}

Our algorithm for Result \ref{res:2} makes use of certain structural properties of the graph for resolving 3-augmenting paths, in addition to employing the maximal matching algorithm of \cite{li2024dynamic} as a subroutine. 
In more detail, we first identify all 3-augmenting paths that traverse two newly inserted edges, as well as ``triangles'' that lead to 3-augmenting paths, and resolve all of these locally by sending the induced subgraph to a player. 
Then, we can focus on 3-augmenting paths that traverse only a single inserted edge, and we point out that there can be a large number of them, which prohibits local processing. 
Instead, we consider a much smaller subgraph, called $G_*'$, which includes  the edges of a certain maximal matching on a bipartite graph.
We show that it is sufficient to break up 3-augmenting paths in $G_*'$ to avoid any 3-augmenting paths in the entire graph.

To give a concrete instance of the upper bound stated in Result~2 for the large-scale data setting, we can set the link bandwidth $\beta$ to be $\Theta(n)$. 
Under this assumption, we can handle batches of $O(\sqrt{n})$ incremental updates in just $O(\log n)$ rounds.

\onlyShort{
We discuss additional related work in the attached full paper.
}
\onlyLong{
\subsection{Additional Related Work}
Most closely related to our work is ~\cite{li2024dynamic}, who study dynamic maximal matching in the $\kclique$ model. 
They give lower bounds and corresponding algorithms for both oblivious as well as adaptive adversaries. 
Notice that a maximal matching is a local problem, in contrast to the more challenging ``global'' problem of computing an (approximate) maximum matching. Dynamic $2/3$-approximate maximum matching was also stuided in the MPC model. The paper \cite{italiano2019dynamic} gave a constant-round dynamic algorithm for $2/3$-approximate maximum matching for single updates, instead of batches of updates. Their idea is to maintain a maximal matching without 3-augmenting path, which is similar to the idea in this paper. We should notice that in this paper, we handle batches of updates, more challenging and complicated. For approximate maximum matching in the MPC model, \cite{ghaffari2018improved} proposes a $(1+\epsilon)$-approximate maximum matching in $O(\log \log n)$ rounds with $\widetilde O(n)$ memory per machine. By simulating their algorithm in the $\kclique$ model, we can obtain a $(1+\epsilon)$-approximate maximum matching in roughly $O(\lceil \frac{m}{k^2 \beta} \rceil +\frac{n}{k\beta})$ rounds if each vertex of the input graph and each of its incident edges assigned to machines are chosen independently and uniformly at random. 

We next discuss related work in the centralized settings. 
There is a survey on recent advances in fully dynamic algorithms \cite{hanauer2022recent}. Edge degree constrained subgraphs (EDCS) \cite{bernstein2015fully} is an elegant and popular data structure to maintain matching for dynamic changes in centralized settings. To the best of our knowledge, the first incremental maximum matching is given by Gupta \cite{gupta2014maintaining}. 
Subsequently, \cite{grandoni2019} shows a deterministic algorithm that maintains a $(1+\varepsilon)$-approximate matching with constant amortized update time per insertion. The techniques of \cite{grandoni2019} are based on augmenting paths elimination. 
Recently, \cite{blikstad2023incremental} further improved the update time complexity for maintaining an incremental $(1-\varepsilon)$-approximate matching to be $O(\poly(1/\varepsilon))$ by weighted EDCS.  
On the lower bound side, it is known that maintaining an exact matching in the incremental/decremental settings require $\Omega(n^{0.333-o(1)})$ update time assuming the $\text{3-SUM}$ conjecture \cite{kopelowitz2016higher}. Even for bipartite graphs, the lower bound for maintaining a maximum matching in the incremental graph needs $\Omega(n^{1-
\varepsilon})$ \cite{dahlgaard2016hardness} under the OMv conjecture.

From the above description of the progress of dynamic (approximate) maximum matching in centralized settings, we can see that EDCS is a crucial tool for dynamic matching. Unfortunately, in most distributed settings, especially for models with small bandwidth, it is hard to maintain an EDCS because of the restricted communication bandwidth.  
For maintaining a $\frac{2}{3}$-approximate maximum matching, a natural idea is to keep that the graph has no augmenting paths of length three. In the MPC model (where the total bandwidth can be $\widetilde O(m)$ where $m$ is the number of edges of the input graph), the algorithm of \cite{italiano2019dynamic} can update a $\frac{2}{3}$-approximate cardinality maximum matching in constant rounds when considering a single update.
Notice that in clique networks with small bandwidth, it would be expensive to achieve this goal if we find such augmenting paths by searching neighborhoods.  While an alternative way is to maintain a 2-hop neighborhood, doing so is prohibitively expensive in the distributed setting.

Approximate maximum matching has been studied extensively the streaming setting. It is straightforward to transform single-pass semi-streaming algorithms to incremental algorithms  in the $\kclique$ model. 
Unfortunately, the approximation ratio obtained from the transformation cannot be as good as $2/3$, as \cite{kapralov2013better} shows that there is no single-pass semi-streaming algorithm for achieving $1-1/e \approx 0.63 < 2/3$ approximation to maximum matching. 
On the other hand, multi-pass streaming algorithms can obtain $(1-\epsilon)$-approximate maximum matching (see the survey~\cite{assadi2023recent}). An interesting direction for future exploration is to transform multiple-pass streaming algorithms to incremental algorithms in the distributed settings. 

When we consider edge insertions/deletions in streams, it is corresponding to dynamic streams (or turnstile streams\footnote{They are nearly the same thing in the literature.}). Dynamic streaming algorithms can be considered as dynamic algorithms with limited space. Due to this property, there are strong relationships between dynamic streaming algorithms and dynamic algorithms in the distributed models. The paper \cite{ahn2012analyzing} initiates the study of dynamic graph streaming algorithms. After that, \cite{assadi2016maximum} shows that the space complexity of single-pass turnstile streaming algorithms for $n^\epsilon$-approximate maximum matching is $\Omega(n^{2-3\epsilon})$, and a similar lower bound for bipartite matching with expected approximation $n^\epsilon$ is also given in \cite{konrad2015maximum}. For random streams, i.e., edges arrive in a random order, ~\cite{farhadi2020approximate} proposes a single-pass semi-streaming algorithm that finds approximately 0.545-approximation of the maximum matching in a general graph.
}

\section{Preliminaries} \label{sec:prelim}

Our algorithms make use of the following routing primitive for the $\kclique$ model, which enables us to efficiently broadcast $N$ tokens to all players, no matter at which players the tokens are located initially:

\begin{lemma}[Algorithm Spreading~\cite{li2024dynamic}] \label{lem:spreading}
Suppose that there are $N \ge \beta(k-1)$ tokens, each of size $\Theta(\log n)$ bits, located at arbitrary players. There exists a deterministic algorithm $\spreading$ such that each player can receive all $N$ tokens in $O\lt(\lceil \frac{N}{\beta\,k} \rceil \rt)$ rounds.
\end{lemma}


For the lower bounds, we use Shannon entropy and mutual information which are defined as follows. 
\begin{definition} \label{def:entropy}
Let $X$ be a random variable. The \emph{entropy of $X$} is defined as
  \begin{align}
    \HH[ X ] = \sum_x \Pr[ X \!=\! x] \log_2(1 /\Pr[ X \!=\! x]). \label{eq:entropy}
  \end{align}
  The \emph{conditional entropy of $X$ conditioned on $Y$} is given by
  \begin{align}\label{eq:conditional_entropy}
    \HH[ X \mid Y ] &= \EE_y[ \HH[ X \mid Y \!=\! y] ] \\
      &= \sum_{y}^{} \Pr[ Y \!=\! y] \HH[ X \mid Y \!=\! y].\notag
  \end{align}
\end{definition}
Note that we have 
\begin{align}
	\HH\lt[ X \rt] \le \log_2\lt( \text{supp}(X) \rt),
\label{eq:uniform}
\end{align}
  where $\text{supp}(X)$ is the support of $X$, and equality is attained in \eqref{eq:uniform} if and only if $X$ is uniformly distributed on $\text{supp}(X)$.

\begin{definition} \label{def:mutual}
  Let $X$, $Y$, and $Z$ be discrete random variables.
  The \emph{conditional mutual information of $X$ and $Y$} is defined as
  \begin{align}
    \II[ X : Y \mid Z ]
      &= \HH[ X \mid Z ] - \HH[ X \mid Y, Z ] \label{eq:mutual_prop2}.
  \end{align}
\end{definition}

\begin{lemma}[Data Processing Inequality, see Theorem 2.8.1 in \cite{cover1999elements}] \label{lem:dataprocessing}
  If random variables $X$, $Y$, and $Z$ form the Markov chain $X \ra Y \ra Z$, i.e., the conditional distribution of $Z$ depends only on $Y$ and is conditionally independent of $X$, then we have
  \[
    \II[ X : Y ] \ge \II[ X : Z ].
  \]
\end{lemma}

\section{A Fully Dynamic Approximate Maximum Matching Algorithm} \label{sec:easy_alg}

In this section, we give an algorithm for handling single edge updates (Result~\ref{res:1}). 

\begin{theorem}\label{thm:simpleAlgo}
There exists a memoryless  $k$-clique algorithm that maintains a maximal matching without 3-augmenting paths when an adaptive adversary inserts or deletes a single edge in each update. 
With high probability, the algorithm sends  $O(\sqrt{m}\log n)$ bits and has an update time complexity of $O(\lceil \frac{\sqrt{m}}{k\beta} \rceil)$ rounds where $m$ is the number of edges in the current graph. 
\end{theorem}

\onlyLong{
\begin{proof}
We first describe how to handle the insertion of a single edge $\set{u,v}$. 
Players $P_u$ and $P_v$ check whether $u$ and $v$ were both matched, in which case no further action is required. 
Similarly, if neither of them was matched, players $P_u$ and $P_v$ simply add the edge $\set{u,v}$ to the matching.
Note that this cannot cause any 3-augmenting paths, because this would require both $u$ and $v$ to have a free neighbor, contradicting the assumption that the matching prior to the update was maximal.
The more challenging case is that only one of the two nodes, say $v$, was matched to some node $w$ whereas $u$ is free: 
Player $P_v$ instructs $P_w$ to check with all neighbors of $w$ whether $w$ has any free neighbor. This requires $P_w$ to send a single message to all players, and each player who has at least one free neighbor $x$ of $w$ (note that this is part of the players' initial knowledge) responds affirmatively by sending $x$'s ID.
If no such $x$ is found, we are again done, as it is straightforward to see that there cannot exist a 3-augmenting path. 
Otherwise, the four players $P_u$, $P_v$, $P_w$, and $P_x$ then coordinate to update the matching by replacing the edge $\set{v,w}$ with the two edges $\set{u,v}$ and $\set{w,x}$, which can be done in $O(1)$ additional rounds. 
Note that this results in all four involved vertices to become matched, and hence this cannot create any new 3-augmenting path in the resulting matching.

Next, we focus on handling the deletion of the edge $\set{u,v}$. 
If $\set{u,v}$ was not in the matching, we do not need to take any further action.
Thus, assume that both $u$ and $v$ become free after removing the matched edge $\set{u,v}$. 
We now describe a procedure that is initiated twice in sequence: once by player $P_u$ and then by $P_v$.

As a first step, player $P_u$ checks with the players hosting $u$'s neighbors whether $u$ has any free neighbor $w$:
\begin{enumerate} 
\item 
If $w$ exists, we add $\set{u,w}$ to the matching.
Note that it is not possible that this causes a new 3-augmenting path, since $w$ was free in the maximal matching prior to the update and hence cannot have any free neighbors.

\item 
We now focus on the more involved case where $u$ itself is free after the deletion, but all of its neighbors are matched. We again consider two cases:
  \begin{enumerate} 
  \item 
 
 Suppose that $u$'s degree $d(u)$ is such that $d(u) \leq  2\sqrt{m}$.  
  Then, $P_u$ uses $\spreading$ (see Lem.~\ref{lem:spreading}) to broadcast the IDs of $u$'s neighbors, denoted by $N(u)$, to all players, which takes at most $O(\lceil \frac{\sqrt{m}}{k\beta}\rceil)$ rounds. 
  As a result, every player $P_{w_i}$, who hosts a vertex $w_i$ that is matched to some $v_i \in N(u)$, knows that $v_i$ is matched to a neighbor of $u$.   
  Next, $P_{w_i}$ checks with the other players whether $w_i$ has a free neighbor.
  We point out that $P_{w_i}$ may be hosting multiple vertices that are matched to (distinct) neighbors of $u$'s, and, in that case, $P_{w_i}$ performs this check via $\spreading$ in parallel for all such vertices.
  By assumption, there are at most $2\sqrt{m}$ such matched vertices for which the players need to perform this check (in parallel), and hence $\spreading$ takes $O(\lceil \frac{\sqrt{m}}{k\beta}\rceil )$ rounds to complete according to Lemma~\ref{lem:spreading}. 
  Any player who hosts at least one free neighbor $x_i$ of $w_i$, picks one of them arbitrarily, and then broadcasts the ID of $x_i$; again, all of these (at most $2\sqrt{m}$) IDs are sent  by the players via $\spreading$.
  Consequently, player $P_u$ learns at least one 3-augmenting path  that $u$ is part of (if one exists). 
  $P_u$ chooses one of these paths, say $(u,v_i,w_i,x_i)$, and then directly coordinates with players $P_{v_i}$, $P_{w_i}$, and $P_{x_i}$ to replace the edge $\set{v_i,w_i}$ with the edges $\set{u,v_i}$ and $\set{w_i,x_i}$ in the matching.
  As this does not cause any vertices to become free that were matched before the edge deletion, this operation cannot create new 3-augmenting paths.

\item 

\begin{figure}[t]
  \centering
\includegraphics[scale=1.0]{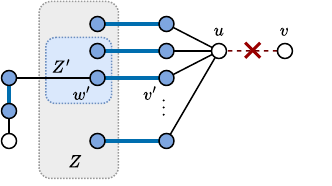}
\caption{\small Case 2(b) in the proof of Theorem~\ref{thm:simpleAlgo}: The edge $\set{u,v}$ is deleted. Blue nodes and blue thick edges correspond to matched nodes and their matched edges. The grey-shaded and blue-shaded areas show the sets $Z$ and $Z'$, respectively. Node $w'$ is the low-degree neighbor found via sampling by $v'$ and, in this example, removing the edge $\set{w',v}$ from the matching will result in an additional 3-augmenting path that needs to be handled accordingly.
}
\label{fig:fully_dynamic_algo}
\end{figure}

Now, suppose that $d(u) > 2\sqrt{m}$: 
Let $Z$ denote the set of nodes that are matched to $u$'s neighbors. 
By a simple counting argument, it follows that there is a subset $Z' \subseteq Z$ of size at least $\frac{|Z|}{2}$, such that every $x \in Z'$ has $d(x) \le 2\sqrt{m}$.
See Figure~\ref{fig:fully_dynamic_algo} for an example.
We describe how $P_u$ can contact the players of $\Gamma = \Theta\lt( \log n \rt)$ randomly sampled vertices in $Z$:
Note that $P_u$ may not know $Z$ (and hence cannot determine the players hosting the vertices in $Z$), but only knows the set of neighbors $N(u)$.
To implement this step, $P_u$ randomly samples $\Gamma$ nodes in $N(u)$ and broadcasts the set of sampled node IDs, denoted by $S$, via $\spreading$ to all players, which takes $O\lt( \lceil \Gamma / \beta k \rceil \rt) = O\lt( \lceil \frac{\log n}{\beta k} \rceil\rt)$ rounds. 
For each sampled $v \in S$ that is matched to some $w \in Z$, player $P_v$ asks $P_w$ to broadcast the ID of $w$ via $\spreading$.
It follows that, with high probability, $P_u$ obtains the node ID of some low-degree vertex $w' \in Z'$ that is matched to some $v' \in N(u)$. (To obtain a Las Vegas algorithm, we can simply repeat this step, if necessary.)
Then, the players $P_u$, $P_{v'}$, and $P_{w'}$ add the edge $\set{u,v'}$ to the matching and unmatch the edge $\set{v',w'}$ instead. 
As a result, there cannot exist any 3-augmenting path traversing $\set{u,w}$, as we assumed that $u$ does not have any free neighbors.
However, this operation may have introduced 3-augmenting paths starting at the now-unmatched node $w'$.

To handle these 3-augmenting paths and ensure maximality, we proceed as follows:
First $P_{w'}$ checks whether $w'$ has any free neighbors and, if yes, we simply add the corresponding edge to the matching. 
Otherwise, we leverage the fact that $w'$ has $d(w') \le 2\sqrt{m}$, which allows us to apply the procedure for Case~(a) above, whereby node $w'$ now takes the role of $u$.

  \end{enumerate}
\end{enumerate}
Finally, we observe the players do not need to keep track of any state across updates apart from the matched edges, which ensures that the algorithm is also memoryless. 
\end{proof}
}

\section{Incremental \texorpdfstring{$\frac{2}{3}-$}-approximate Maximum Matching under Batch-Dynamic Updates} \label{sec:inc}
In this section, we consider an adaptive adversary who can insert a batch of edges all at once.

From Lemma~\ref{lem:hopcroft}, we can see that, for any $\frac{2}{3}$-approximate maximum matching, there is no augmenting path with length $3$, which exists if and only if there is a matched edge whose both endpoints have free neighbors. 
Throughout this section, we restrict ourselves to \emph{incremental updates}, i.e., batches of edge insertions.
The following theorem states our main result for batch dynamic algorithms, which we prove in the remainder of this section.

\begin{theorem}\label{thm:incre-maxm}
    For $\ell$ edge insertions in a batch, there exists an algorithm that can maintain an incremental  $\frac{2}{3}$-approximate maximum matching by maintaining a maximal matching without 3-augmenting paths in $\widetilde O(\frac{\ell}{\sqrt{k\beta}})$ rounds. 
\end{theorem}

\subsection{Description and Analysis of the Algorithm}
Our algorithm splits the batch of $\ell$ new edges into \emph{mini-batches} of $\mathbf{b}=\lfloor\sqrt{k\beta}\rfloor$ edges each, and then it processes these mini-batches sequentially, such that the resulting matching after each mini-batch is maximal and does not contain 3-augmenting paths.
Thus, we focus on describing the algorithm for just a single mini-batch  $B$.
Without loss of generality, we assume that no edges in $B$ connect two matched nodes; we show that the algorithm does not unmatch any currently matched vertices and hence we can simply ignore such edges if they do happen to exist. 

We first introduce some notation. We use $M_0$ to denote the matching before the algorithm performs any updates for mini-batch $B$. We say that an edge is \emph{new} if it is part of $B$, and \emph{old} otherwise.
Similarly, if an edge was inserted between $u$ and a free node $v$, we say that $v$ is a \emph{new free neighbor} of $u$. 
Note that $v$ may be a new free neighbor for one node and simultaneously an old free neighbor for another.

The main algorithm consists of three conceptual phases: 

\subsection*{Phase~1}
Initially, we want to ensure that all players know $B$, and thus we broadcast the entire set to all players using procedure $\spreading$ (see Lem.~\ref{lem:spreading}).
Throughout Phase~1, we focus on a subgraph $G_1$ that is small enough (i.e., consists of $O(\sqrt{k\beta})$ edges) to be collected and processed efficiently by just a single player. 
The edges forming $G_1$ consist of several subsets that we describe next:
\begin{itemize} 
\item 
We define $F \subseteq B$ to be the set of new edges that were inserted between free nodes.
\item 
Let $I_1 \subseteq M$ be the set of (old) matched edges, such that, for every edge $\set{w,u} \in I_1$, there are incident new edges $\set{w,a}$ and $\set{u,b} \in B$ and $a \ne b$.
We use $\hat{I}_1$ to denote the set of edges in $I_1$ together with all of their incident new edges. 

\item We will also process certain triangles in Phase~1 that may be part of 3-augmenting paths:
To this end, consider any triangle $t=\set{w,u,c}$, for some vertices $u$, $w$, and $c$.
We say that $t$ is a \emph{harmful triangle}, if the edge $\set{w,u}$ is matched, $c$ is free, and at least one of $w$ or $u$ has an additional incident new edge to a free neighbor.
Note that this includes the case that at least one of the other two unmatched edges of the triangle (i.e., $\set{w,c}$ or $\set{u,c}$) are new.
Also note that there exists at most one triangle $\{c,w,u\}$ for a matched edge $\{w,u\} \in I_1$ before edge insertions (otherwise, there would have been a 3-augmenting path prior to processing $B$). The harmful triangles also contain such triangles.
On the other hand, we say that $t$ is a \emph{harmless triangle} if it is not harmful. 
Let $T$ be the set of edges forming harmful triangles,  including all incident new free edges in $B$. 
\end{itemize}

\begin{figure}[t]
  \centering
\includegraphics[scale=1.0]{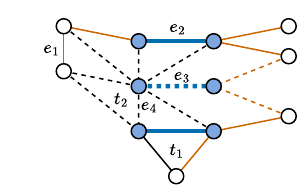}
\caption{The graph $G_1$ used in Phase~1: Newly inserted edges are orange and matched edges are blue. A solid line indicates that the edge is in $G_1$, whereas a dashed line corresponds to a (possibly matched or new) edge that exists but is \emph{not} included in $G_1$. In this example, $e_1 \in F$, $e_2 \in I_1$, $e_3 \notin I_1$. Note that $t_1$ forms a harmful triangle and thus we include its edges and its incident new edges in $G_1$. On the other hand, $t_2$ is omitted, as it is a harmless triangle due to $e_4$ not being matched.}
\label{fig:ph1}
\end{figure}

\begin{lemma} \label{lem:sizeG1}
Let graph $G_1$ be the subgraph consisting of the edges in $F \cup \hat{I}_1 \cup T$. It holds that $|E(G_1)|=O\lt( \sqrt{k\beta} \rt)$.
\end{lemma}
\onlyLong{
\begin{proof}
The lemma follows because the size of $F$, $T$, and $\hat{I}_1$ is bounded by the number of new edges, i.e., $O\lt( \sqrt{k\beta} \rt)$.
\end{proof}
}

Our analysis crucially relies on the invariant that there are no $3$-augmenting paths in subgraph $G_1$, which we guarantee by sending all of these edges to $P_1$, including the status of each edge (i.e., whether it is matched). 
Upon receiving $G_1$, player $P_1$ extends the current matching on $G_1$ to a (locally-computed) maximal matching $M'$, fixes some ordering of the matched edges, and then iteratively removes any 3-augmenting path $p=(a,b,c,d)$, where $\set{b,c}$ is currently matched, by unmatching $\set{b,c}$ and instead matching $\set{a,b}$ and $\set{c,d}$.
Subsequently, all other players update the state of their vertices and edges by learning the computed matching. 
Both of these communication steps are done by invoking Algorithm~$\spreading$. 
We use $M_1$ to denote the updated matching on the whole graph.

\begin{lemma} \label{lem:part1}
  At the end of Phase~1, the following properties hold for the updated matching~$M_1$:
  \begin{enumerate} 
  \item[(a)] Every vertex matched in $M_0$ remains matched in $M_1$. 
  \item[(b)] $M_1$ is maximal with respect to the subgraph $G_1$, i.e., there exist no edges between free nodes in $G_1$.
  \item[(c)] There are no 3-augmenting paths in $G_1$.
  \item[(d)] All triangles in $G$ are harmless.  
  \end{enumerate}
\end{lemma}
\onlyLong{
\begin{proof}
Properties (a)-(c) are immediate from the fact that we compute a maximal matching on $G_1$ and then iteratively resolve all 3-augmenting paths.
For (d), it is sufficient to observe that all harmful triangles (and their incident new edges) are part of $T \subseteq G_1$.
Moreover, the algorithm does not create any new harmful triangles, due to Property~(a).
\end{proof}
}
 
\begin{definition} \label{def:important}
Let $I \subseteq M_1$ be the set of matched edges that have exactly one endpoint with incident new edges connecting to free nodes, whereas the other endpoint is connected to at least one free node via an old edge and does not have any new incident edges to free nodes. 
More precisely, for every edge $\set{w_i,u_i} \in I$, the following hold:
\begin{itemize}
\item[(a)] both $w_i$ and $u_i$ each have at least one distinct free neighbor; 
\item[(b)] $w_i$ has at least one free old neighbor;
\item[(c)] $u_i$ has at least one free new neighbor.
\end{itemize} 
\end{definition}
Before proceeding, we need to define several node sets: 
We use $W$ to denote the set of all $w_i$ that satisfy (b) and are an endpoint of some edge in $I$, and we define $U$ to contain all endpoints of edges in $I$ satisfying (c).
Moreover, we define $V$ to be the set containing every node that is a new free neighbor of some $u_i \in U$.
Finally, set $X$ contains every free node that is an old free neighbor of some node in $W$. 

We first show some properties of these sets that we leverage in the analysis below:

\begin{lemma} \label{lem:X_and_V}
The nodes in $V \cup X$ form an independent set in $G$.
\end{lemma}
\onlyLong{
\begin{proof}
Assume towards a contradiction that there is an edge $e$ between two vertices in $V \cup X$.
Since both endpoints of $e$ are free vertices at the start of Phase~2, and we do not unmatch any previously matched vertex in Phase~1 (see Lem.~\ref{lem:part1}(a)), it follows that that the endpoints of $e$ were free also at the start of Phase~1.
Thus, $e$ must be a new edge, as otherwise the matching that we started with was not maximal.
Since $e \in F$, it must have been included in graph $G_1$, on which we have computed a maximal matching in Phase~1, which is a contradiction.
\end{proof}
}

\begin{lemma} \label{lem:W_U}
The nodes in $W$ do not have new free neighbors, and the nodes in $U$ do not have any old free neighbors.
\end{lemma}
\onlyLong{
\begin{proof}
We first show the result for nodes in $W$.
Assume towards a contradiction that $w_i \in W$ has a new free neighbor $v$. 
Recall that $w_i$ is matched to $u_i$ who also has a new free neighbor $v'$. 
Moreover, by the definition of $I$, node $w_i$ also has an old free neighbor $x \in X$ and it holds that $x \ne v'$.
If $v=v'$, then $(w_i,u_i,v')$ forms a harmful triangle $t$, as there is a 3-augmenting path $p=(x,w_i,u_i,v')$. 
By the definition of graph $G_1$ (see Lem.~\ref{lem:sizeG1}), we know that $p$ is part of $G_1$, contradicting Lemma~\ref{lem:part1}(d).
Now, suppose that $v\ne v'$.
It follows that there exists a 3-augmenting path $(v,w_i,u_i,v')$ that is part of $\hat{I}_1 \subseteq G_1$, because both $\set{v,w_i}$ as well as $\set{u_i,v}$ are new free edges.
This contradicts Lemma~\ref{lem:part1}(c).

Now, assume towards a contradiction that some node $u_i \in U$ has a free old neighbor $x$.
Let $x'$ be the free old neighbor of $w_i$. 
In the case that $x\ne x'$, there exists a 3-augmenting path $p=(x',w_i,u_i,x)$.
By Lemma~\ref{lem:part1}(a) we did not unmatch any vertices in Phase~1, and thus it follows that $p$ also existed in the matching at the start of the update, which contradicts that the previous matching did not contain 3-augmenting paths.
On the other hand, if $x = x'$, we know that $u_i$ must also have a free new neighbor $v$.
It follows that $(x,w_i,u_i)$ forms a harmful triangle, because there is a 3-augmenting path $(x,w_i,u_i,v)$, thus contradicting Lemma~\ref{lem:part1}(d).

\end{proof}
}

\begin{lemma} \label{lem:all_in_I}
Consider the matching $M_1$ resulting from Phase~1. Every 3-augmenting path in $G$ must traverse a matched edge in $I$.
\end{lemma}
\onlyLong{
\begin{proof} 
Suppose that there is a 3-augmenting path $(a,b,c,d)$, where the matched edge is $\set{b,c} \notin I$.
It is not possible that both $a$ and $b$ have a new incident edge, because then $(a,b,c,d)$ would have been part of $\hat{I}_1 \subseteq G_1$ and processed in Phase~1.
Furthermore, it is also not possible that both $a$ and $b$ have an incident old free edge, as, in that case, the matching that we started with prior to the update had a 3-augmenting path, contradicting our assumption.
Thus, it follows that either $\set{a,b}$ or $\set{c,d}$ must be a new edge, which allows us to conclude that $\set{b,c} \in I$. 
\end{proof}
}

\subsection*{Phase~2}
\begin{figure}[t]
\begin{subfigure}[t]{0.3\textwidth}
\centering
\includegraphics[scale=0.9]{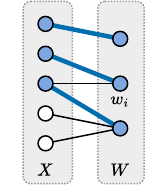}
\caption{Graph $G_{X,W}$}
\label{fig:g_xw}
\end{subfigure}
\hspace{2mm}
\begin{subfigure}[t]{0.3\textwidth}
\includegraphics[scale=0.9]{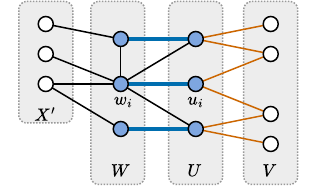}
\caption{Graph $G_*'$}
\label{fig:g_star_dash}
\end{subfigure}
\hspace{2mm}
\begin{subfigure}[t]{0.3\textwidth}
\includegraphics[scale=0.9]{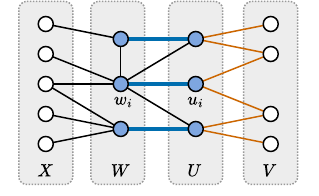}
\caption{Graph $G_*$}
\label{fig:g_star}
\end{subfigure}
  \centering
\caption{The subgraphs on which the algorithm operates in Phases 2 and 3. Thick blue edges are matched and orange edges are part of the batch of inserted edges. Free nodes are colored white.}
\label{fig:gs}
\end{figure}

We use the edge set $I$ (see Def.~\ref{def:important}) to define a new bipartite graph $G_{X,W}$, on which we will operate in Phase~2:
The vertices of $G_{X,W}$ consist of the sets $X$ and $W$. 
We include all edges from $G$ in $G_{X,W}$ that are of the form $\set{x,w}$, where $x \in X$ and $w \in W$, which ensures that $G_{X,W}$ is bipartite. 
In particular, we omit all edges between nodes in $W$ (recall that there cannot be any edges at this point between nodes in $X$ as all of them are free, see Lemma~\ref{lem:X_and_V}.)
Figure~\ref{fig:g_xw} shows an example of $G_{X,W}$.

Next, we make use of the result of \cite{li2024dynamic}, namely that there exists an algorithm that maintains a maximal matching on a graph with a small update time, when the adaptive adversary deletes a batch edges.
A useful consequence of their result is that it yields a fast algorithm for computing a maximal matching on bipartite graphs, as we show next:

\begin{lemma} \label{lem:itcs}
Consider an $n$-node bipartite graph $H$ with vertex sets $A$ and $B$ that is partitioned in a balanced way among $k$ players, and assume that each vertex is marked as being either in $A$ or $B$.
There exists an algorithm $\EdgeDel$ that computes a maximal matching on $H$ in $O\lt( \frac{s\log n}{\sqrt{\beta k}} \rt)$ rounds with high probability, where $s = \min(|A|,|B|)$. 
\end{lemma}
\onlyLong{
\begin{proof}
Without loss of generality, assume that $|A| \le |B|$.
Since each player knows which of its vertices are in $A$ and which ones are in $B$, we can simply aggregate these counts at player $P_1$ and ensure that every player learns that $ s = |A|$.
Then, for each node $a \in A$, the player $P_a$ locally adds a neighbor $a'$ to $H$ and pretends that the edge $\set{a,a'}$ is matched, which ensures that there is a maximal matching on this extended graph.
Note that the resulting graph partition is still balanced.  
Then, we remove all $\set{a,a'}$ edges, to which the players respond by executing the update algorithm of \cite{li2024dynamic}. 
Since the number of deleted edges is $s$, Theorem~1.2 in \cite{li2024dynamic} tells us that this algorithm completes in $O\lt( \frac{s\log(n)}{\sqrt{\beta k}} \rt)$ rounds, as required. 
\end{proof}

}

Equipped with Lemma~\ref{lem:itcs}, we can compute a maximal matching $M_{X,W}$ on $G_{X,W}$, and we define $X'\subseteq X$ to be the subset of ``left'' nodes that were matched to some of the nodes in $W$ in $M_{X,W}$.
For each node $w_i \in W$, we use $x_i \in X'$ to denote the matched node in $M_{X,W}$ (if any).
Since $|W|=O\lt( \sqrt{k\beta} \rt)$, computing $M_{X,W}$ takes $O(\log n)$ rounds by executing $\EdgeDel$ on $G_{X,W}$.
We say that $M_{X,W}$ is a \emph{virtual matching}, as we do not update the actual matching on the graph at this point.

To complete Phase~2, we ensure that player $P_1$ learns $M_{X,W}$ (which includes $X'$) via $\spreading$.
Since $|M_{X,W}|=O\lt( \sqrt{k\beta} \rt)$, this requires only $O(1)$ rounds, and thus it follows that the time complexity of Phase~2 is dominated by the execution of $\EdgeDel$:
\begin{lemma} \label{lem:part2}
Phase~2 has a running time of $O\lt( \log n \rt)$ rounds with high probability.
\end{lemma}

\subsection*{Phase~3}
In the final phase, we operate on the subgraph $G_*'=G[X' \cup W \cup U \cup V]$.
By definition, every edge in $I$ has at least one incident new edge (at the start of Phase~2), and the same is true for the vertices in $V$.
This means that the number of vertices in $G_*'$ is $O(\sqrt{k\beta})$, and thus the number of edges (new or old) in $G_*'$ is $O({k\beta})$. 
We send all edges of $G_*'$ to player $P_1$ in $O(1)$ rounds for local processing via $\spreading$ (see Lem.~\ref{lem:spreading}).
Starting from matching $M_1$, the local computation at $P_1$ produces the final matching $M$, by processing each $\set{w_i,u_i} \in I$ in sequence, thereby iteratively augmenting the matching obtained in Phase~1. (Recall that, in Phase~2, we only computed a virtual matching without modifying the currently matched edges.) 
Note that we cannot simply instruct $P_1$ to locally compute a maximum matching on $G_*'$, as we require the resulting matching to have certain properties, which we formally state in Lemma~\ref{lem:p3} below.

In more detail, we fix some order of the edges in $I$ and proceed in two stages:
\begin{description} 
\item[Stage 1:] 
For each edge $\set{w_i,u_i} \in I$ ($i\ge 1$), player $P_1$ checks whether there is a 3-augmenting path $p=(x_i,w_i,u_i,v)$ in $G_*'$, for any free $v \in V$, where $x_i \in X'$ denotes the vertex to which $w_i$ was matched in $M_{X,W}$, if it exists.
If there is such a path, then $P_1$ updates $M$ by replacing $\set{w_i,u_i}$ with the edges $\set{x_i,w_i}$ and $\set{u_i,v}$ in the matching, before moving on to edge $\set{w_{i+1},u_{i+1}}$.
\item[Stage 2:] 
Next, player $P_1$ inspects every (currently) matched edge $\set{w_i,u_i} \in I$ and checks whether there exists any 3-augmenting path for any free $x\in X'$ and $v \in V$; note that it is not necessarily the case that $x=x_i$. If such a path exists, then we again resolve it by replacing $\set{w_i,u_i}$ with the two other incident edges in the 3-augmenting path.
After all edges in $I$ have been processed in the above manner, $P_1$ sends the final matching $M$ to every player via $\spreading$.
\end{description}

For our analysis, we define graph $G_* = G[X \cup W \cup U \cup V]$.
Note that $G_*\supseteq G_*'$, and $G_*$ may in fact contain up to $\Theta\lt( n \rt)$ nodes, which prohibits us from sending the entire graph to just a single player. 

\begin{lemma} \label{lem:p3}
The following properties hold for Phase~3: 
\begin{enumerate} 
\item[(A)] The algorithm does not unmatch any node in $X\cup W \cup U \cup V$ that was matched at the end of Phase~1.
Moreover, if a previously free node is matched at some point in Phase~3, it remains matched until the end of the algorithm.
\item[(B)] For every edge $\set{w_i,u_i} \in I$, it holds that either
 \begin{itemize} 
 \item $\set{w_i,u_i}$ remains matched, or 
 \item $w_i$ is matched to some node $x \in X'$, whereas $u_i$ is matched to some node in $V$. 
 Moreover, if $x \ne x_i$, i.e., $w_i$ was not matched to the same node that it was matched to in the virtual matching $M_{X,W}$, then $x_i$ is not free in $M$.
 \end{itemize}
\item[(C)] $M$ is a maximal matching on $G_*$.
\end{enumerate}
\end{lemma}

\onlyLong{
\begin{proof} 
Property~(A) is immediate from the description of the algorithm, since the local operations performed at player $P_1$ do not unmatch any currently matched vertices.
For (B), it is sufficient to observe that Stage~1 ensures $\set{x_i,w_i}$ is matched if,  at the point when $\set{w_i,u_i}$ is processed, there is a 3-augmenting path across $\set{w_i,u_i}$ and $x_i$ is free. 
It remains to prove (C): The crucial observation is that, given the matching at the start of Phase~3, there are no free edges in $G_*$. 
To see why this is true, it is sufficient to recall that every edge in $I$ is matched and that Lemma~\ref{lem:X_and_V} ensures that $X \cup V$ forms an independent set. 
\end{proof}
}

\begin{lemma} \label{lem:nofree}
At the end of the algorithm, the nodes in subgraph $G_*$ do not have any free neighbors in $G \setminus  G_*$.
\end{lemma}

\onlyLong{

\begin{proof}
Assume towards a contradiction that there is a node $a \in G_*$ who has a free neighbor $b \in G \setminus G_*$.
We distinguish several cases.
If $a \in X \cup V$, then $a$ is free at the start of Phase~3. According to the algorithm, we do not unmatch any matched nodes in Phases~1 and 2 (see Lemma~\ref{lem:part1}(a)), and thus it follows that the edge $\set{a,b}$ must have been unmatched and both $a$ and $b$ must have been free at the start of Phase~1. 
Note that $\set{a,b}$ cannot be an old edge, as the initial matching $M_0$ was maximal.
Thus, $\set{a,b}$ is new, which means that it will be part of set $F \subseteq G_1$. 
Due to Lemma~\ref{lem:part1}, we know that Phase~1 computes a maximum matching on $G_1$, which contradicts the fact that $a$ and $b$ were free after Phase~1.

Now, consider the case $a \in W$, and recall that all old free neighbors of $a$ at the start of Phase~3 are in the set $X$, and hence part of graph $G_*$. 
Moreover, Lemma~\ref{lem:W_U} tells us that $a$ does not have any new free neighbors. This shows that $b$ does not exist. 
Similarly, in the case where $a \in U$, it follows again by Lemma~\ref{lem:W_U} that $a$ does not have any old free neighbors, and we know that every new free neighbor of $a$ is in $V$, and thus part of $G_*$.
\end{proof}
}

\begin{lemma} \label{lem:gstar}
The matching $M$ obtained at the end of Phase~3 is maximal in $G_*$ and does not admit any 3-augmenting paths in $G_*$.
In addition, all vertices that were matched at the end of Phase~1 remain matched. 
\end{lemma}

\onlyLong{
\begin{proof}
From the description of the algorithm, it is clear that the operations performed by player $P_1$ do not unmatch any vertices when resolving 3-augmenting paths, which proves the second part of the lemma.

Assume towards a contradiction that there is a 3-augmenting path $p$.
We distinguish several cases, depending on the matched edge $e$ that $p$ traverses. 
 
By Lemma~\ref{lem:X_and_V}, we know that $X$ and $V$ form an independent set, which rules out that $e$ has both of its endpoints in $X \cup V$.
Recall that, at the start of Phase~3, all matched edges in $G_*$ were the edges in $I$, which means that any matched edge is of the type $\set{w_i,u_i}$, where $w_i \in W$ and $u_i \in U$.  
In conjunction with Lemma~\ref{lem:p3}(B), it follows that the only possibility for a matched edge $e$ in $G_*$ is one of the following cases:
\begin{enumerate} 
\item[(1)] $e \in I$; 
\item[(2)] $e$ has one endpoint in $X$ and the other in $W$;
\item[(3)] $e$ has one endpoint in $U$ and the other in $V$.
\end{enumerate}

Consider Case (1), i.e., $p=(a,w_i,u_i,b)$ such that $\set{w_i,u_i} \in I$.
Since $a$ and $b$ are free, Lemma~\ref{lem:nofree} tells us that $a,b \in G_*$. 
Moreover, by Lemmas~\ref{lem:p3}(B) and \ref{lem:W_U}, we know that $a \in X$ and $b \in V$.
Since Stage~2 removes all 3-augmenting paths across $\set{w_i,u_i}$ in the graph $G_*'$, but $p$ exists, it follows that $a \in X \setminus  X'$.
Moreover, we know that there does not exist any edge $\set{x_i,w_i}$ in the virtual matching $M_{X,W}$, as we did not match $w_i$ to $x_i$ in Stage~1, which means that $w_i$ was unmatched in $M_{X,W}$.
Thus, it is possible to add the edge $\set{a,w_i}$ to $M_{X, W}$, which, however, contradicts the maximality of $M_{X,W}$ guaranteed by Lemma~\ref{lem:itcs}.

Next, we consider Case (2), i.e.,  the matched edge of $p$ is $e = \set{x,w_i}$, where $x \in X$ and $w_i \in W$. 
Then, by Lemmas~\ref{lem:X_and_V} and \ref{lem:W_U}, it follows that the first (free) node in $p$ must be in $W$, which, however contradicts Lemma~\ref{lem:p3}(A), since all nodes in $W$ were matched.
To complete the proof, observe that the argument for Case (3) is analogous. 
\end{proof}
}

\begin{lemma} \label{lem:correct}
The final matching $M$ is maximal and there are no 3-augmenting paths in $G$. 
\end{lemma}
\onlyLong{
\begin{proof}
Assume towards a contradiction that there exists a 3-augmenting path $p=(a,b,c,d)$ in $G$.
By combining Lemmas~\ref{lem:gstar} and \ref{lem:nofree}, we know that the matched edge $\set{b,c}$ cannot be in $G_*$, and in particular $\set{b,c} \notin I$, which means that it is not possible for $\set{b,c}$ to be incident to exactly one new edge connecting to a free node.
Clearly, $\set{a,b}$ and $\set{c,d}$ cannot both be free edges, as otherwise, we would have resolved the resulting 3-augmenting path already in Phase~1.
Thus, the only choice is that $\set{a,b}$ and $\set{c,d}$ are both old edges.
Since matched nodes are not unmatched at any point in the algorithm (see Lem.~\ref{lem:part1}(a) and \ref{lem:p3}(A)), it follows that path $p$ already must have existed at the start of Phase~1, contradicting the assumption that the previous matching did not have any 3-augmenting paths.
\end{proof}
}
\begin{lemma} \label{lem:updatetime}
The running time of Phases~1-3 is $O\lt(\log n \rt)$ rounds with high probability.
\end{lemma}
\onlyLong{
\begin{proof} 
In Phases~1 and 3, all communication performed by the algorithm consists of various invocations of $\spreading$ with at most $O({k\beta})$ tokens. By  Lemma~\ref{lem:spreading}, each of these calls completes in $O\lt(\log n\rt)$ rounds. 
Moreover, Phase~2 takes $O(\log n)$ rounds according to Lemma~\ref{lem:part2}.
\end{proof}
}

To complete the proof of Theorem~\ref{thm:incre-maxm}, recall that the running time in Lemma~\ref{lem:updatetime} holds for a single mini-batch of $\Theta\lt( \sqrt{k\beta} \rt)$ edge insertions. 
By taking a union bound over the $\Theta\lt( \ell / \sqrt{k\beta} \rt)$ mini-batches, it follows that Lemmas~\ref{lem:correct} and \ref{lem:updatetime} also hold for the entire batch $B$, and we obtain the claimed update time of $O\lt( \frac{\ell \log n}{\sqrt{k\beta}} \rt)$.

\subsection{The Lower Bound under Edge Insertions} \label{sec:lbs}

In this section, we investigate the fundamental limitations on the update time achievable in the $\kclique$ model. We give a lower bound for any adaptive adversary who not only determines the initial (balanced) vertex partitioning but also has access to the states of nodes including the obtained approximate maximum matching throughout the execution.

{
\begin{theorem} \label{thm:lb_insertions}
Suppose that every matched edge $e$ is output by every player who hosts an endpoint of $e$.
Any $\kclique$ algorithm that, with high probability, maintains a maximal matching without 3-augmenting paths under batches of $\ell$ edge-insertions against an adaptive adversary, has update time complexity of $\Omega\lt( \frac{\ell}{\beta k\log n} \rt)$ rounds. 
\end{theorem}

}
We use a similar lower bound construction as in the proof of Theorem~1.1 of \cite{li2024dynamic}.
Consider a graph $G$ that consists of $n/5$ segments $L_1,\ldots,L_q$, whereby each $L_i$ consists of five nodes $t_i,u_i,v_i,w_i,x_i$ and will contain two edges at first: $\{u_i,v_i\}$ and $\{v_i,w_i\}$.\footnote{For simplicity, we assume that $n/5$ and $n/k$ are integers, as this does not impact the asymptotic bounds.} We say that $v_i$ is the \emph{middle} vertex of $L_i$.
If we conceptually place the line segments on a $5 \times (n/5)$ grid, we can tile the grid using $k$ ($1\times(n/5k)$)-rectangles $R_1,\ldots,R_k$, to cover all nodes (without overlaps), whereby each rectangle covers exactly $n/5k$ vertices. 
In more detail, we restrict the choice of the rectangles such that every player holds the vertices of exactly $5$ of these rectangles with the restrictions that the player obtains at most one vertex from $\{u_i,v_i,w_i\}$, for each index $i\in[n/5]$.
Note that the vertex assignment is possible since we are assuming an adaptive adversary.

The adversary first inserts the edges $\{u_i,v_i\}$ and $\{v_i,w_i\}$ sequentially,  for every $i\in[n/5]$. It is easy to see that in the resulting graph, also denoted by $G$, any maximal matching is a maximum matching. 
Moreover, there is a player $P$ who hosts a set $S_P$ of at least $n/(5k)$ middle vertices $v_i$. Notice that the adaptive adversary knows $P$ and $S_P$ when selecting the edge insertions. Let $I_P$ be the set of indexes of the line segments, for which there is a node $v_i \in S_P$, and note that $\lvert I_P \rvert \geq n/(5k)$.

The adversary samples a subset $J_P \subseteq I_P$ of size $\ell$ u.a.r., and also samples u.a.r.\ an $\ell$-length vector $\Gamma$ of random bits.
Then, for each $i \in J_P$, the adversary inserts either the edge $\{t_i,u_i\}$ if $\Gamma[i]=0$ (Case 1) or, if $\Gamma[i]=1$, it inserts the edge $\{w_i,x_i\}$ (Case 2).
The intuition behind the hardness of the lower bound is that player $P$ hosting the set $S_P$ has no way of distinguishing between Case 1 and Case 2 without further communication with the other players. Observe that, before inserting the $\ell$ randomly chose edges in the line segments $L_i$ $(i \in [J_P])$, either $\{u_i,v_i\}$ or $\{v_i,w_i\}$ must be in the matching. Without loss of generality, assume that $\{u_i,v_i\}$ is matched. 
Since the player $P$ does not host any vertex incident to an inserted edge, it remains unaware of the edge insertions unless it communicates with other players. 
It is easy to verify that, the status of the edges $\set{u_i,v_i}$ and $\set{v_i,w_i}$ must change (from matched to unmatched and vice versa) to avoid a 3-augmenting path, if and only if the adversary chooses to insert the edge $\set{t_i,u_i}$ for line segment $L_i$, which happens with probability $\frac{1}{2}$.

\textbf{Proof of Theorem~\ref{thm:lb_insertions}}.
By a slight abuse of notation, we also use $P$ as a random variable that captures $P$'s current state, i.e., hosted vertices and currently matched incident edges. We provide more power to the algorithm by revealing $J_P$ to player $P$ for free. 
Given $J_P$, there are $2^\ell$ equally-likely choices for the vector $\Gamma$, and thus
\begin{align}
   \HH\lt[ \Gamma \ \md|\ J_P,P \rt] 
   =  \Omega(\ell)\label{eq:lb1}.
\end{align}

Let random variable $\mathsf{Out}$ be the edges that are output by $P$ as a part of a matching without 3-augmenting paths after the algorithm's response to the batch of $\ell$ edge insertions. 
We use $\textbf{1}_{\text{Succ}}$ to denote the indicator random variable that the algorithm succeeds, and recall that it is $1$ with high probability (i.e., with probability at least $1-1/n$).

\begin{lemma}\label{lem:entropy}
    $\HH[\Gamma \mid \mathsf{Out}, J_P,P] = O(1)$.
\end{lemma}
\onlyLong{
\begin{proof}
 By the chain rule of conditional entropy, we have that 
\begin{align*}
  \HH [\Gamma \mid \textsf{Out},  J_P,P] &\leq \HH [ \Gamma , \mathbf{1}_{\text{Succ}} \mid \textsf{Out},  J_P,P] \\
  &= \HH[\Gamma \mid \mathbf{1}_{\text{Succ}} \textsf{Out},  J_P,P] + \HH[\mathbf{1}_{\text{Succ}} \mid \textsf{Out},  J_P,P] \\
  &\leq \HH[ \Gamma \mid \textbf{1}_{\text{Succ}}, \textsf{Out},  J_P,P] + 1 \\
  &= \mathbb{P}[\textbf{1}_{\text{Succ}} = 1] \HH[ \Gamma \mid \textsf{Out},  J_P,P, \textbf{1}_{\text{Succ}} = 1] \\ 
  &\phantom{----}+\mathbb{P}[\textbf{1}_{\text{Succ}} = 0] \HH[ \Gamma \mid \textsf{Out},  J_P,P, \textbf{1}_{\text{Succ}} = 0]+1 \\
  &\leq \HH [ \Gamma  \mid \textsf{Out},  J_P,P, \textsf{1}_{\text{Succ}}=1] + \frac{1}{n}\HH[ \Gamma  \mid \textbf{1}_{\text{Succ}}=0]+1 \\
  &= \HH[ \Gamma  \mid \textsf{Out},  J_P,P, \textsf{1}_{\text{Succ}}=1] + O(1),
\end{align*}
where the final inequality follows because $|\Gamma| = \ell \le n$.

As we have conditioned on the event $\textbf{1}_{\text{Succ}}=1$, player $P$ must (correctly) decide how to update the states of edges incident to $v_i$.
If $\set{u_i,v_i}$ was matched before the update, and the algorithm's output matches $\set{v_i,w_i}$ instead, then $P$ knows that the adversary must have inserted $\set{t_i,v_i}$, and thus $\Gamma[i]=0$.
A similar statement  holds when $\set{v_i,w_i}$ was matched before the update.
Thus, player $P$ can deduce the entire vector $\Gamma$ from the algorithm's output, which implies that 
$\HH[\Gamma \mid \textsf{Out}, J_P,P, \textsf{1}_{\text{Succ}}=1]  = 0$.
\end{proof}
}
Now, we can show the lower bound. 
Let $\Pi$ be the transcript of messages received by player $P$ during the edge insertions. We should notice that $\mathsf{Out}$ is a function of $\Pi$ and the local state of $P$ including the public randomness. Given $P$, we have a Markov Chain: $(P,J_P) \to \Pi \to \mathsf{Out}$. By the data-processing inequality, Lemma~\ref{lem:entropy} and \eqref{eq:lb1}, it follows that
\begin{align*}
    \begin{aligned} 
    		\HH[\Pi] \ge
        \II[\Pi:\Gamma \mid J_P,P] \geq \II [\mathsf{Out}: \Gamma \mid J_P,P] 
        = \HH[\Gamma  \mid J_P,P] - \HH[J_P \mid \mathsf{Out},J_P,P]
        \geq \Omega(\ell) - O(1) 
        = \Omega(\ell).
    \end{aligned}
\end{align*}

$\HH[\Pi]$ is maximized when $\Pi$ is uniformly distributed, i.e., $\HH[\Pi]\leq \log_2 2^{c\beta (k-1)T\log n+1}$, where $T$ denotes the number of rounds that the algorithm runs. Thus, we have
$
c\beta(k-1)T\log n + 1 \geq \HH[\Pi] \geq \HH[\Pi \mid P] -\HH[\Pi \mid J_P, P] 
=  \II[\Pi:J_P\mid P] = \Omega(\ell),
$
and we conclude that $T = \Omega(\frac{\ell}{\beta k \log n})$.

\begin{acks}
\end{acks}

\bibliographystyle{ACM-Reference-Format}
\bibliography{ref}

\end{document}